\documentclass{aastex}
\usepackage{amsmath,amssymb}
\begin{document}
\title{Neutron  Star  Superfluidity,\\Dynamics  and  Precession}
\author{M. Ali Alpar}
\affil{Sabanc{\i} University\\
Orhanl{\i}, Tuzla, 34956 Istanbul, Turkey}
\email{alpar@sabanciuniv.edu}

\begin{abstract}
Basic rotational and magnetic properties of neutron superfluids and 
proton superconductors in neutron stars are reviewed. The modes of 
precession of the neutron superfluid are discussed in detail. We emphasize 
that at finite temperature, pinning of superfluid vortices does not offer 
any constraint on the precession. Any pinning energies can be 
surmounted by thermal activation and there exists a dynamical steady state in 
which the superfluid follows the precession of the crust at a small lag 
angle between the crust and superfluid rotation velocity vectors. At 
this small lag the system is far from the critical conditions for 
unpinning, even if the observed precession of the crust may entail a large 
angle between the figure axis and the crust's rotation velocity vector. We 
conclude that if long period modulations of pulse arrival times and 
pulse shapes observed in a pulsar like the PSR B1828-11 are due to the 
precession of the neutron star, this does not have any binding implications 
about the existence of pinning by flux lines or the existence of 
Type II superconductivity in the neutron star.  

\end{abstract}

\section{Introduction}
This lecture consists of two parts.  The first part is a general review of neutron superfluidity and proton superconductivity, 
and the determining effects of superfluidity and superconductivity on neutron star dynamics. The second part is a 
discussion of neutron star precession and possible constraints imposed by superfluidity on precession, with a critical 
review of recent inferences on the nature of proton superconductivity in 
neutron stars based on the occurrence of precession.  In Marmaris I devoted roughly half of the lecture to the first part. In the written version here 
I will summarize  this general discussion, and supplement it with references to lectures on neutron star superfluidity and 
superconductivity in previous ASI of this series (Sauls, 1989; Pines, 1991; Alpar, 1991, 1995, 1998, 2001). 
Taken together, those lectures form a fairly comprehensive introduction 
to the subject. The bulk of the exposition here will be devoted to a discussion of precession, on the lines and level of detail of the 
original lecture. 

\section{Superfluidity and Superconductivity in Neutron Stars}

Like all interacting fermion systems at low temperature, neutrons and protons in the neutron star 
interior are expected to be superfluid. Neutron superfluidity exists in two different regimes. 
In the inner crust regions of the neutron star, neutrons are found in bound states in the nuclei 
that form a solid lattice in the crust, as well as the continuum Bloch states of the crust lattice. 
The neutrons in the continuum states form a superfluid that coexists with the lattice. This 
coexistence determines the particular dynamical properties of the neutron star. The neutron and 
proton superfluids in the core of the neutron star exist in a homogeneous medium where their 
dynamical properties are very different from those of the crust superfluid. 

The minimum free energy state of any rotating system is the state of rigid body rotation. 
A charged superfluid - a superconductor - like the protons in the neutron star core achieves 
the state of rigid body rotation by setting up a magnetic field called the London field, of magnitude 
$B_{London}= (2mc/e) \Omega$ sustained by supercurrents at the boundaries of the superconducting region. 
The energy cost of the London field is negligible compared to the free energy gain of the rigid rotation 
state. A London field of only 10$^{ - 4}$ G is needed for neutron star rotation rates $\Omega \sim$ 100 rad s$^{ - 1}$. 

The neutron superfluid achieves a minimum energy state very close to the state of rigid body rotation by 
forming an array of quantized vortices, each carrying a quantum of vorticity 
$\kappa = h/2m_n$ where {\em h} is the Planck 
constant and $m_{n}$ is the neutron mass. A uniform area density of vortices 
$n_0$ = 2$\Omega /\kappa$ sets up a 
macroscopic rotation state of the superfluid that is the state of rigid 
rotation almost everywhere, except for 
the vicinity of individual vortex lines. The cores of the vortex lines, 
within a distance $ \xi  \sim E_F/ (k_F \Delta) $ 
from the vortex axis, are populated with neutrons in the normal phase. Here $E_F$ and $k_F$ are the Fermi 
energy and wavenumber of the neutrons and $\Delta$ is the neutron superfluid's energy gap.  
It is the interaction of these vortex cores with the normal matter around them 
that conveys to the superfluid any differential rotation with respect to the normal matter components of the star, including the outer crust.
The motion of the vortices in response changes the rotational state of the superfluid to follow the spinup or spindown of the 
outer crust and normal matter under the external torques on the star. In the core superfluid the neutron vortices interact with a 
continuous, homogeneous system of electrons and protons. The dynamics of the neutron superfluid in the core of the neutron star is 
determined by the vortex-electron interaction, which is expected to be very efficient (Alpar \& Sauls 1988) on both observational and theoretical grounds. 
This strong coupling results from a spontaneous magnetization of the neutron
vortex lines (Alpar, Langer \& Sauls 1984) and is expected to 
achieve rotational equilibrium 
between the neutron superfluid and the normal matter including the outer crust of the star with a time lag $\tau \sim$ 400 P only, where P 
is the rotation period of the neutron star. 

In the inner crust vortex lines can get pinned to the nuclei forming the crust lattice. Pinning will constrain the rotation rate of the superfluid, 
and leads to a lag between the rotation rate of the superfluid and the normal matter outer crust of the star as the normal matter spins down 
(or up) in response to external torques. The vortex lines will then have an average flow rate through thermal activation against the 
pinning energy barriers (Alpar et al. 1984a,b, Alpar, Cheng \& Pines 1989). By this thermal creep mechanism the pinned crustal superfluid can also follow the spindown (or up) of the 
star in response to external torques. In fact, in the rotational equilibrium state for the pinned crust superfluid, as well as 
for the homogeneous core superfluid, the superfluid and the normal matter components spin down (or up) at the same rate. 
Observations of glitching pulsars indicate that these neutron stars are always close to their rotational equilibrium state. 

Magnetic free energy is minimized in the state with a uniform magnetic field distributing the total flux throughout the body of the system. 
For most observed stars the required magnetic field that corresponds to the dipole magnetic moment of the star is of the order of 
10$^{ 12}$ G, in the direction of the magnetic moment. (The magnetic moment 
constraint is distinct from the rotational constraint 
which, for the protons, is met by the tiny London field.) The superconducting protons can get very close to the free energy minimum by sustaining 
a uniform array of flux lines each carrying a quantum of flux $\Phi_0 = hc / 2e$ at an area density $n_f$ 
corresponding to the required 
uniform field B. This density of flux lines is $ n_f $= B / $\Phi_0$. The central structure of the flux lines contain normal matter 
within a length scale $\xi$ and magnetic field within a length scale $\lambda \equiv c (4\pi n_p e^2/m_p)^{-1/2}$ called the ''London penetration depth''. 
Here $c$ is the velocity of light, $n_p, m_p$ and $e$ are the proton number density, mass and charge respectively. 
As is the case with the neutron vortex lines, 
there is an energy cost associated with the core of each flux line. Wherever a flux line and a vortex line intersect, the junction 
formed will represent an energy gain because of the overlap of the flux line core with the vortex line core. This means the 
two lines are pinned to each other at the junction. Any change in the rotational state of the neutron superfluid has to involve a motion 
of the vortex lines, which is not possible, topologically, without encountering junctions with flux lines and therefore being hindered by 
pinning energy barriers. This should have important consequences for the short term rotational behaviour as well for the rotational and magnetic 
evolution of the star. Ruderman, Zhu \& Chen (1998) have argued that the pulsar spindown will proceed with the motion of the vortex lines 
either carrying pinned 
magnetic flux lines with them, in the case of pulsars of the age of the Vela pulsar ($\sim$ 10$^4$ yrs) or older, or just cutting through the flux lines 
in the case of  younger pulsars like the Crab pulsar.

\section{Precession}

The precession of neutron stars first attracted attention in connection with the 35 d 
cycle of Her X-1  which is likely to involve the precession of the accretion disk which might be coupled to the precession of 
the neutron star. The involvement in precession of the pinned superfluid in the neutron star crust was discussed 
in some detail. Shaham (1977) pointed out that pinning will keep the angular momentum of the pinned superfluid fixed in 
the body frame (the rotating and precessing frame of the neutron star's solid crust), and that such a constraint will 
require $\omega_{pr}/\Omega \sim$ I$_p /$I. In the case of pinning the precession frequency would be determined
by the ratio of pinned moment of inertia I$_p$ to total moment of inertia. 
Alpar \& \"{O}gelman (1987) showed that this constraint will be avoided in practice because 
at finite temperature vortex creep can provide a steady state in which  the rotation velocity vector $\Omega_p$
of the pinned superfluid follows the precession of the rotation velocity vector $\Omega_c$ of the 
crust. The vector lag $\Omega_c - \Omega_p$ that drives the precession by creep is perpendicular to 
$\Omega_p$, and has a small magnitude. Alpar \& \"{O}gelman (1987) also showed that precession by vortex creep has a 
steady state only in the linear regime of vortex creep, with a lag small compared to $\omega_{cr}$ for unpinning. (The  
nonlinear regime of vortex creep in a spinning down pulsar, in which the lag in steady state is close to but slightly smaller than 
$\omega_{cr}$, does not have an analogue for precession by creep.) Periods and 
amplitudes that are consistent with precession driven by the expected effective triaxialities for neutron 
stars do not necessarily rule out the presence of pinning.  In short, at finite temperatures pinning constraints are 
not absolute. Furthermore, when steady state creep allows the pinned superfluid 
components to take part in precession, the steady state lag is much smaller than the lag needed for unpinning, 
and spontaneous unpinning is unlikely.  These early discussions investigated the pinned crust superfluid, 
and possibility of pinning in the core superfluid at vortex line-flux line junctions was ignored. However, 
these arguments are equally applicable to any site of pinning.  

In recent years the issue of neutron star precession has attracted renewed attention as a result of 
observations of what seems to be free precession in the radio pulsar PSR B1828-11 (Stairs, Lyne \& Shemar 2000, Hobbs et al. 2004). 
The evidence comes as 
correlated periodic changes in the pulse shape and arrival times of the pulses at the rotation 
period of the radio pulsar. This is naturally explained in terms of the pulsar beam nodding 
about the line of sight as the neutron star precesses. Another line of explanation could invoke the periodic 
migrations of magnetic field patterns rooted on the neutron star surface, roughly in analogy with the model for the bursting, accreting pulsar 
GRO J1744-28 (Miller 1996). Here we will discuss neutron star precession, and application of precession models to the case of PSR B1828-11.

An important pinning constraint would 
arise from the pinning of neutron vortex lines in the core of the neutron star by their junctions 
with the proton flux lines. The importance of this interaction for neutron star dynamics was first pointed out by Sauls (1989). In view of the fact that the potentially pinned core neutron superfluid 
comprises most of the neutron star's moment of inertia, 
Shaham's absolute pinning argument would require a precession period that is comparable 
to the rotation period, which is not the case observed. The amplitude angles of precession required to fit the observed wandering of pulse arrival times 
are larger than the critical angle between the rotation axis of the pinned superfluid and that of the crust 
and normal matter under absolute pinning. Vortex lines would unpin, but this would not free the superfluid from the absolute pinning constraint 
as the vortex lines would encounter new pinning junctions with flux lines. This process is highly dissipative. The observed amplitudes of precession 
would not be sustained by an absolutely pinned superfluid if the angle or rotation rate difference between the crust, together with its pinned lines, 
and the superfluid is of the order of $\Omega$ times the observed precession amplitude right from the beginning of the precession of the crust. 

This  argument, relying on absolute pinning, was invoked to conclude that the 
core neutron superfluid is not pinned, and therefore that the proton superconductor does not contain quantized 
flux tubes that would pin the neutron superfluid's vortex lines. Thus it is claimed that the proton superconductor in 
neutron stars is Type I and not Type II (Link 2003). In a Type I superconductor magnetic free energy is minimized 
by alternating macroscopic layers of magnetized normal matter and diamagnetic regions where the protons are superconducting. 
This conclusion, if definite, would have important consequences. 
Calculations of the proton superconducting energy gap at the high density neutron star interior are highly uncertain 
because of our incomplete knowledge of the strong interaction and because of the difficulties of the many body calculations 
at these supernuclear densities. Evidence for Type I rather than Type II superconductivity would reverse the constraint 
on the proton superconductivity gap $\Delta_p  (\rho)$ provided by the condition $\lambda$ < (>) $\xi / \sqrt{2}$ for Type I (Type II) 
superconductivity. Thus $\Delta_p  (MeV)$, the value of the proton energy gap in units of MeV, would be less than ( greater than) 
0.09 (x/0.05)$^{5/6}  (\rho_{14})^{5/6}$ for Type I (Type II) superconductivity - here x is the ratio of the number density of protons 
to the number density of nucleons, and $\rho_{14}$ is the density in units of 10$^{14}$ gm cm$^{-3}$. More importantly for astrophysics, 
the existence of pinning between vortex and flux lines has as yet not fully explored consequences for neutron star dynamics and evolution. 
Particularly interesting is the idea that flux vortex pinning would induce magnetic field decay coupled to the spindown 
of the neutron star on evolutionary timescales, providing an intriguing qualitative explanation for the prevalence of low magnetic fields among 
old neutron stars in low mass X-ray binaries and millisecond pulsars (Srinivasan et al. 1990).  This attractive possibility is not viable 
if neutron stars have Type I proton superconductors. But the entire argument leading to these sweeping conclusions 
relies on the prevalence of absolute pinning, as would hold only at absolute zero temperature. The allowance of precession 
by vortex creep means a loophole in the argument. Jones (2004) has already commented that consideration of pinning and energy dissipation in the 
{\em crust} superfluid shows that any precession would be damped. A careful discussion by Alpar \& \"{O}gelman (1987) showed that vortex creep 
in the crust superfluid and dissipative vortex electron coupling in the core superfluid make precession possible, but also provide strong damping. 
Here we extend the discussion of Alpar \& \"{O}gelman (1987) to the consequences of pinning of neutron vortices in the core superfluid 
at junctions with the proton flux lines. 

Let us now examine the conditions for precession by creep more carefully. We shall take a two component model for the neutron star. 
The component which is coupled directly to the external torque $\vec{N}_{ext}$ consists of the crust, normal matter including the electrons in the core 
and the proton superconductor together with its flux lines. 
The other component, which contains most of the moment of inertia, is the neutron superfluid in the core of the star. The two components are coupled 
through an internal torque $\vec{N}_{int}$. We label the two components with $c$ for crust and $n$ for the neutron superfluid in the core. 
The Euler equations describing the motion of the rotation vectors $\Omega_c$ and $\Omega_n$ are:
\begin{eqnarray}
I_c\frac{d\vec{\Omega}_c }{dt} + \vec{\Omega}_c  \times \vec{L}_c & = & \vec{N}_{ext} + \vec
{N}_{int} \\
I_n\frac{d\vec{\Omega}_n }{dt} + \vec{\Omega}_c  \times \vec{L}_n & = & - \vec{N}_{int} 
\end{eqnarray}
where $\vec{L}_c$ and  $\vec{L}_n$ are the angular momenta of the two components. 

Let us note, as a generality, that the precession of any {\em fluid} components of a body are driven by internal torques which depend on rotational velocity lags 
between the solid and the fluid. (We shall discuss specific models of this velocity dependence below.)
With velocity dependent torques, energy will be dissipated and the precession of the fluid will inevitably be damped.

In the case of strict pinning of the vortex lines the angular momentum of the superfluid in the body (crust, $c$ ) would remain constant. 
Adding the Euler equations for the two components, with the condition 
\begin{eqnarray}
I_n\frac{d\vec{\Omega}_n }{dt} & = & \frac{d\vec{L}_n}{dt} = 0, \\
I_c\frac{d\vec{\Omega}_c }{dt} +\vec{\Omega}_c  \times (\vec{L}_c + \vec{L}_n) & = & \vec{N}_{ext}
\end{eqnarray}
is obtained as the equation governing the precession of the neutron star with pinned superfluid. The case of free precession is obtained by 
setting $ \vec{N}_{ext} $ = 0. As suggested by the form of Eq.(4), the precession frequency under pinning is of the order of $ (I_n + I_c) /I_c $ 
times the rotation frequency $\Omega$. The solution of Eq.(4) gives precession at the frequency:
\begin{equation}
\omega_{pr} =  \epsilon \Omega + \frac{I_n }{I_c} \Omega 
\end{equation}
for any precession angle. The distinction between $\Omega_c$ and $\Omega_n$ is negligible here. 

If the core neutron superfluid is absolutely pinned to the crust system, which includes the charged proton superconductor, as a result of the 
absolute pinning of the neutron vortex lines to the proton flux lines, Eqs. (4) and (5) would apply.  
As the ratio of the core neutron superfluid and the crust (including all normal matter and charged components ) moments of inertia 
is of the order of 10 or maybe more the inferred precession 
of PSR B 1828-11, with $ \omega_{pr} \cong$ 7.7 $\times$ 10$^{-8}$ clearly cannot be 
supported by neutron stars with absolutely pinned core superfluids, or for that matter, with pinned crust superfluids (Alpar and \"{O}gelman 1987). 

At finite temperature a superfluid can change its rotational state by vortex creep even in the presence of pinning. As noted above the 
theory of vortex creep was developed first in the context of pulsar spindown, glitches and postglitch response. The work of Alpar and \"{O}gelman (1987) 
applied the vortex creep model to the precession of the inner crust superfluid in which the vortices are pinned to the nuclei that form the crust lattice. 
The core superfluid was treated in terms of the dynamical coupling that arises from the scattering electrons 
( and charge-coupled superconducting protons) by the spontaneously magnetized neutron vortex cores. 
The magnitude of the damping torque for the core superfluid, 
and therefore the magnitude of the external counter-torque that would be needed to keep the apparently free precession 
behaviour of Her X-1 in its 35 d cycle were estimated, and found to be commensurate with the torques available from the 
accretion disk, though the vector matching, and locking in, of the external torque at the right phase to keep driving precession 
might be difficult. It is now considered more likely that the 35 d cycle of Her X-1 is actually due to the modulation of the 
accretion column and the emerging beam of radiation by the precession of the accretion disk, not necessarily entailing 
any precession of the neutron star. The contribution of the Alpar and \"{O}gelman paper is the demonstration that analogously 
with the situation for spindown (or spinup) vortex creep enables a pinned superfluid to take part in precession too. 

Let us now write the Euler equations for our two component model, taking into account only the core superfluid, which contains 
most of the moment of inertia of the star
and modeling the internal torque that couples the pinned superfluid to the crust as due to the angular momentum transfer through vortex creep: 
\begin{equation}
\frac{d\vec{\Omega}_n }{dt} + \vec{\Omega}_c  \times \vec{\Omega}_n = - \frac{\vec{N}_{int}}{ I_n} = \frac{\vec{\Omega}_c - \vec{\Omega}_n }{\tau}
\end{equation}
Here we have taken the moment of inertia I$_n$ to be isotropic. Alpar and \"{O}gelman have shown that 
steady state creep for precession has to be in the linear regime: The second term on the left hand side, $\vec{\Omega}_c  \times \vec{\Omega}_n$, is 
perpendicular to the $\vec{\Omega}_c  \vec{\Omega}_n$ plane, while the torque $\vec{N}_{int}$ has to be in that plane. Therefore 
$\vec{\Omega}_c  \times \vec{\Omega}_n$ must be balanced by a component of the first term $d\vec{\Omega}_n / dt$. In the nonlinear creep regime 
steady state the magnitude $|\vec{\Omega}_c  \times \vec{\Omega}_n| \sim  \omega_{cr} \Omega$, which is much larger than the 
magnitude of the first term, 
$|d\vec{\Omega}_n / dt| \sim \omega_{pr} \Omega$. Hence Eq.(6) does not have a steady state creep solution in the nonlinear regime. 
The form of the internal torque is therefore given as appropriate for the linear regime in the last equality in Eq. (6). Thus the torque is proportional to 
$(\vec{\Omega}_c - \vec{\Omega}_n)/\tau$ where $\tau$ is the linear creep regime relaxation time. The component of this equation parallel to 
$\vec{\Omega}_n$  describes spindown or spinup, while the component perpendicular to $\vec{\Omega}_n$ is of interest here, as it describes 
precession:
\begin{equation}
\Omega_n \frac{d\hat{\Omega}_n }{dt} + \vec{\Omega}_c  \times \vec{\Omega}_n =  \frac{(\vec{\Omega}_c - \vec{\Omega}_n)_{\perp}}{\tau}.
\end{equation}
The subscript $_{\perp}$ means  "perpendicular to $\vec{\Omega}_n$", and 
$\hat{\Omega}_n$  is the unit vector for $\vec{\Omega}_n$. In a steady state the rotation vector $\vec{\Omega}_n (t)$ for the pinned core superfluid 
would follow the precessing rotation velocity vector $(\vec{\Omega}_c (t)$ of the crust in the body frame with a fixed angular relation. We try the solutions 
\begin{eqnarray}
\vec{\Omega}_c(t) & = &  \Omega_c  [ \sin \alpha \cos(\omega_{pr} t ) ,  
\sin \alpha \sin (\omega_{pr} t ) , \cos \alpha ] \\
\vec{\Omega}_n (t) & = &  \Omega_n  [ \sin \beta \cos(\omega_{pr} t - \phi_n) ,  
\sin \beta \sin (\omega_{pr} t - \phi_n) , \cos \beta ]
\end{eqnarray}
In these trial solutions we have taken the precession of the crust as given. The precession frequency $\omega_{pr}$ of the crust is set by the 
crust's effective 
triaxiality.  The rotation vectors $\vec{\Omega}_c$ and $\vec{\Omega}_n$ precess with respective cone 
angles $\alpha$ and $\beta \equiv \alpha - \epsilon$ 
around the crust figure axis, as shown in Fig. 1. 
\begin{figure}[ht]
\begin{center}
\includegraphics[width=10.0cm]{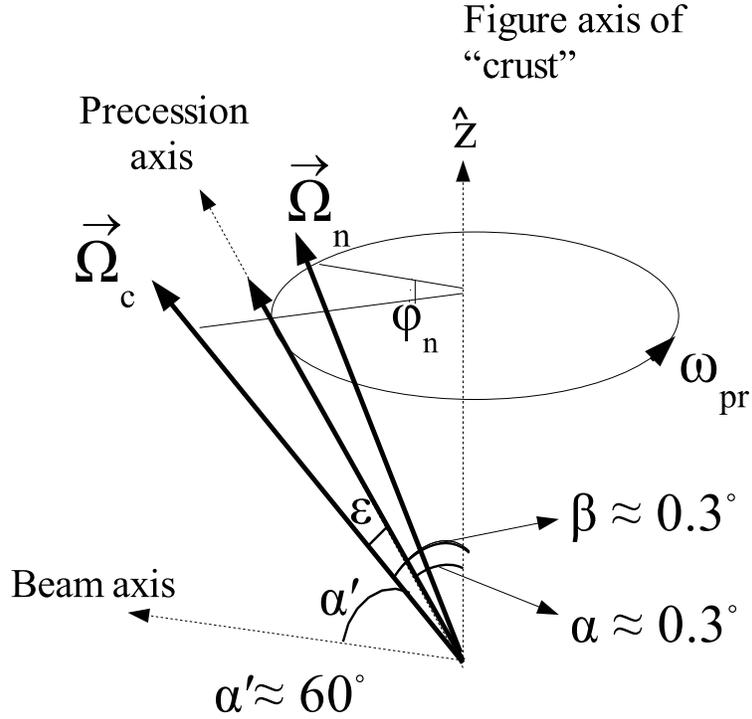}
\end{center}
\caption{Precession of PSR B1828-11. In the corotating frame of the crust precession is around the figure axis $\hat{z}$. In the inertial frame precession 
takes place around the Precession axis, which is coplanar with $\hat{z}$ and $\vec{\Omega}_c$. }
\end{figure}

The angle $\phi_n$ is the lag with which the pinned superfluid's angular velocity follows behind 
that of the crust in precession. We find that a solution for steady state precession of the fluid component exists, with 
\begin{eqnarray}
\epsilon & \cong & \frac{\omega_{pr} \sin \alpha}{\Omega} \\
\phi_n & \cong &  \frac{\omega_{pr}}{\Omega^2 \tau} << \epsilon.
\end{eqnarray}
From its relation with $\omega_{pr}$ the angle $\epsilon$ is seen to be commensurate with an effective triaxiality.

We now check if such a solution can be achieved with the physical parameters of the vortex creep model. 
This solution is indeed in the linear creep regime since the lag required for creep by precession is much less than the critical lag for unpinning, 
$|\Omega_c - \Omega_n| \sim \epsilon \Omega << \omega_{cr}$. Even though the wobble angle $\alpha \sim $ 0.3$^{o}$ itself could exceed 
likely unpinning values, it is the relative velocity between the pinned vortex lines and the ambient superfluid, therefore the 
lag $|\Omega_c - \Omega_n|$, that should exceed $\omega_{cr}$ for unpinning. Thus in the linear regime steady state precession by creep is very 
unlikely to reach unpinning conditions. 

The precession rate 
\begin{equation}
\dot{\Omega}_{n, \perp} = \omega_{pr} \Omega_{n, \perp}
\end{equation}
is sustained by creep if the average creep velocity $<v>_{\perp}$  in the direction of precession  satisfies 
\begin{eqnarray}
\omega_{pr} & = & \frac{<v>_{\perp}}{r}  \nonumber  \\
& = & \frac{v_ 0}{r} \exp {( - E_p/kT)}\ 2\ \sinh (\frac{ E_p |(\vec{\Omega}_c - \vec{\Omega}_n)_{\perp}|}{ kT \omega_{cr}}) \nonumber \\ 
& \cong & 2 \frac{v_ 0}{r} \frac{ E_p |(\vec{\Omega}_c - \vec{\Omega}_n)_{\perp}|}{ kT \omega_{cr}}  \exp {( - E_p/kT)}.
\end{eqnarray}
Here r denotes the distance from the rotation axis, E$_p$ denotes the pinning energy at each vortex line-flux line junction and v$_0$ 
an average microscopic fluctuation velocity of the vortex lines (Alpar et al. 1984a). The geometry is as shown in Fig. 2b of Alpar and \"{O}gelman (1987). 
The last equality is valid in the linear regime. Using Eqns (12 ) and ( 13), we write  
\begin{equation}
\dot{\Omega}_{n, \perp} = \frac{(\vec{\Omega}_c - \vec{\Omega}_n)_{\perp}}{\tau_{\perp}}
\end{equation}
where the timescale for the linear response of precession to the lag $ {(\Omega_c - \Omega_n)}_{\perp}$ is
\begin{equation}
\tau_{\perp} = \frac{r}{2 v_ 0} \frac{ kT \omega_{cr}} { E_p \Omega_{n, \perp}} 
\exp (E_p/kT) = \frac{ kT } { 2 v_ 0 \rho \kappa \lambda \ell_f \Omega_{n, \perp}} \exp(E_p/kT).
\end{equation}  
Here the ratio between the critical frequency for unpinning $\omega_{cr}$ and the pinning energy E$_p$ has been expressed in 
terms of the superfluid density $\rho$, the vortex quantum $\kappa$, the length scale across a pinning junction, which in the 
present case is the London length $\lambda$ and the distance between successive pinning junctions along a vortex line, which 
is simply the distance $\ell_f$ between flux lines. 

In steady state precession, using Eqs. (13) and (14),
\begin{equation}
|(\vec{\Omega}_c - \vec{\Omega}_n)_{\perp}| = \omega_{pr} \Omega\ \sin \alpha\ \tau_{\perp}.
\end{equation}

The requirement that the creep by precession is actually in the linear regime means that the argument of the sinh in Eq. (14) must be less than 1, 
\begin{equation}
\frac{ E_p  |(\vec{\Omega}_c - \vec{\Omega}_n)_{\perp}|}{ kT \omega_{cr}} =  \frac{ E_p  \omega_{pr}  \Omega\ \sin \alpha\ \tau_{\perp}}{ kT \omega_{cr}} < 1.
\end{equation}
From Eqs. (16) and (18) we obtain
\begin{equation}
E_p  <  kT \ln (\frac{ 2 v_ 0 }{r \omega_{pr}}) 
\end{equation}

To estimate the temperature {\em kT} in the neutron star core we start with an estimate of the surface temperature $T_s$. 
The dominant contribution to energy dissipation in the neutron star is due to the energy dissipation in vortex creep for the spindown 
of the pinned crust superfluid. Energy dissipation rates in all other dynamical couplings between the normal matter and the pinned 
crust and core superfluids, including spindown or precession by creep are negligible in comparison. This can be easily verified by 
calculating the energy dissipation rate $ \dot{E}_{diss}$ for each process. For neutron stars past their initial cooling stage, the surface 
temperature can be estimated by equating the energy dissipation rate from the spindown of the pinned crust superfluid  to the blackbody 
luminosity of the neutron star surface (Alpar et al. 1984a),
\begin{equation}
\dot{E}_{diss} = I_p \omega_{cr} |\dot{\Omega}| = 4\pi R^2 \sigma {T_s}^4 . 
\end{equation}
Here $I_p \sim$ 10$^{43}$ gm cm$^2$ is the moment of inertia of the pinned crust superfluid. $\omega_{cr} \leq$ 1 rad s$^{-1}$ is the 
estimated steady state lag for spindown by creep, which should be prevalently in the nonlinear regime for this pulsar (Alpar, Cheng \& Pines 1989). 
On the right hand side, {\em R} is the neutron star radius and $\sigma$ is the Stefan-Boltzmann constant. For PSR B1828-11, we 
estimate the surface temperature as $T_s$ = 5 $\times$ 10$^4$ K. Using the Gudmunsson, Pethick and Epstein (1982) relation 
between the surface temperature and the core temperature of a neutron star, we estimate the core temperature as {\em kT} $\cong$ 2.3 keV. 

Using v$_ 0 \sim$ 10$^6$ cm s$^{-1}$, and r$ \sim$10$^6$ cm in Eq.(19), we find that  the typical pinning energy E$_p$ at each vortex line-flux 
line junction must be less than about 40 keV in order for precession by vortex creep, which, as mentioned above, has to be in the linear regime 
to be possible at the estimated temperature of PSR B1828-11. This is substantially less than earlier estimates of E$_p \sim$ 1 MeV. However, 
there are many uncertainties in E$_p$. There is no real calculation, but qualitatively, effects of tension in vortex and flux lines, of small junction 
angles achieved by bending and of collective effects involving small angle pinning of a vortex line to many flux lines making up a very weakly 
defined junction, analogous to superweak pinning in the case of vortex-lattice interactions in the crust (Alpar et al. 1984b), all point towards weaker pinning. 
A particularly important consideration is the fact that the core neutron superfluid is dynamically coupled to the crust and normal matter electrons with a short coupling time of the order of only 400 times the rotation period. The proton superconductor and the flux lines anchored in it must also follow the motion of the electrons and therefore the crust because of very tight electromagnetic coupling. The flux lines and vortex lines will therefore have reduced relative velocities and effective pinning energies - the electromagnetic forces tend to enforce corotation of the two line systems and will therefore reduce the pinning energies. We therefore conclude that a steady state of precession in the presence of pinning might well be possible for the core superfluid in the neutron star.   

The linear creep regime has a steady state lag that is much less than the critical lag for unpinning, so that if the steady state precession by creep can be attained the pinned superfluid will be far from unpinning conditions. The precession of the crust may have been triggered by a rare event, like a glitch, that offset the pinned superfluid and the crust rotation rates to a level comparable to unpinning conditions. In that case the initial response of creep in the pinned superfluid will be in the nonlinear regime, which does not have a steady state. This initial response will be in the form of rapid vortex creep which will bring the lag down, closer to the steady state in the linear regime. The timescale to relax into steady state will be of the order of the linear regime relaxation time $\tau_{\perp} $. For the steady state to be reached within one period of the crust precession, we have the condition: 
\begin{equation}
\tau_{\perp} \leq P_{pr} \sim 1000 \; {\rm days},  
\end{equation}
for PSR B1828-11. This translates into the requirement 
\begin{equation}
E_p \cong 47 keV + 2.3 keV(1/2 \ln(\rho_{14}/ B_{12}))
\end{equation}
at the estimated core temperature of 2.3 keV, using $\Omega $ 16 rad s$^{-1}$  for PSR B1828-11 and appropriate values of $\lambda \cong $ and of 
$\ell_f$ normalized for $\rho $ = 10$^{14}$ g cm$^{-3}$ and {\em B} = 10$^{12}$ G. We already found that steady state precession by creep is possible if 
typical pinning energies E$_p$ are less than about 38 keV. The condition for the existence of steady state precession by creep satisfies the condition to reach such a steady state within one precession period. 

Once precession is set up in an isolated neutron star like PSR B1828-11, for how long can it survive? Jones (2004) has noted that it would take only a small fraction of the pinning energy per junction to be dissipated in each precession period for precession to be damped. Thus, the observed timing and pulse shape excursions might be due to overdamped precession in the presence of pinning, and the observation of such a transient would not imply that pinning and Type II superconductivity do not exist. Let us now take a careful look at the damping of precession in terms of the coupling mechanisms between the pinned core superfluid and the crust. Fluids will damp precession as they can freely adjuct their shape. The timescale $\tau_{pr}$ for damping precession is expected to satisfy the Bondi-Gold relation
\begin{equation}
\omega_{pr} \tau_{pr} \sim \Omega \tau_{rot}  
\end{equation} 
where $\tau_{rot}$ is the dynamical coupling or damping time between the rotation rates of the solid and fluid components of the star. For the core neutron superfluid, the shortest rotational coupling/damping time is not due to the vortex line-flux line interactions, but rather due to the vortex line-electron interactions. As we mentioned in Section 1, the core neutron superfluid is dynamically coupled to the crust and normal matter through the interactions of the spontaneously magnetized neutron vortex lines and the charged particles (electrons and superconducting protons which are electromagnetically very tightly coupled with the electrons). This coupling time is expected to be $\tau_{rot} \sim$ 400 rotation periods from theory (Alpar \& Sauls 1988). This is supported by observational bounds from the Vela pulsar glitches (see Dodson, McCulloch \& Lewis for the Vela glitch with the tightest observational resolution). 
The associated precession damping time is
\begin{equation}
\tau_{pr} \sim 400 P_{pr}  \sim 400 yrs
\end{equation}
for PSR B1828-11. While this is only a small fraction of the pulsar's lifetime, given the total amount of observation time that can resolve 
precessional excursions devoted to the entire pulsar sample, and if occasional fraction of the glitches that are expected to occur in every pulsar do trigger precession, the probability of our observing precession from some (one or two) single pulsars might not be all that small. Using models developed for the Vela pulsar glitches in a manner consistent with the statistics of all pulsar glitches, 
the time between glitches in PSR B1828-11 is estimated to be about 80 yrs. Thus if one in five glitches of PSR B1828-11 resets the effective 
triaxiality of the solid crust to trigger precession observable by us, the pulsar would be sustained in extended precession for intervals of more 
than one damping time. 

\section*{Summary}
In summary, we find that neutron stars can precess in the presence of vortex line-flux line pinning constraints, 
so that precession does not necessarily show that pinning and Type II superconductivity do not exist in neutron stars. 
Furthermore, even the tightest damping mechanism we know, based on observational constraints from pulsar spindown and glitches, 
do not make it completely unlikely that we observe some single pulsars in precession.

\section*{Acknowledgments}
This work was supported by Sabanc{\i} University
Astrophysics and Space Forum and by the Turkish Academy of Sciences. I thank Altan Baykal, Sinan Kaan Yerli,  \c{C}a\u{g}da\c{s} \.{I}nam and the members of the LOC for their help, and 
\"{U}nal Ertan for the figure.

\end{document}